 \documentclass[iop]{emulateapj}
\shorttitle{Chromospheric evaporation in the 29 March 2014 flare}
\shortauthors{M. Battaglia et al.}
\begin{document}
\renewcommand{\labelitemi}{-}
\title{How important are electron beams in driving chromospheric evaporation in the 2014 March 29 flare? }
\author{Marina Battaglia\altaffilmark{1}, Lucia Kleint \altaffilmark{1}, S\"am Krucker \altaffilmark{1,2}, David Graham \altaffilmark{3} }
\affil{Institute of 4D Technologies, School of Engineering, University of Applied Sciences and Arts Northwestern Switzerland, CH-5210 Windisch, Switzerland} 
\affil{Space Sciences Laboratory, University of California, Berkeley, CA 94720-7450, USA} 
\affil{	INAF-Osservatorio Astrofisico di Arcetri, Firenze, Italy}
\email{marina.battaglia@fhnw.ch}
\begin{abstract}
We present high spatial resolution observations of chromospheric evaporation in the flare SOL2014-03-29T17:48. Interface Region Imaging Spectrograph (IRIS)  observations of the \ion {Fe}{21} $\lambda 1354.1$ line indicate evaporating plasma at a temperature of 10 MK along the flare ribbon during the flare peak and several minutes into the decay phase with upflow velocities between 30 km s$^{-1}$ and 200 km s$^{-1}$. Hard X-ray (HXR) footpoints were observed by RHESSI for two minutes during the peak of the flare. Their locations coincided with the locations of the upflows in parts of the southern flare ribbon but the HXR footpoint source preceded the observation of upflows in \ion {Fe}{21} by 30--75 seconds. 
However, in other parts of the southern ribbon and in the northern ribbon the observed upflows were not coincident with a HXR source in time nor space, most prominently during the decay phase. In this case evaporation is likely caused by energy input via a conductive flux that is established between the hot (25 MK) coronal source, which is present during the whole observed time-interval, and the chromosphere. 
The presented observations suggest that conduction may drive evaporation not only during the decay phase but also during the flare peak. Electron beam heating may only play a role in driving evaporation during the initial phases of the flare.

\end{abstract}
\maketitle
\section{INTRODUCTION} \label{Introduction}
In the standard solar flare model beams of accelerated electrons propagate from the acceleration site in the corona to the dense chromosphere where they deposit their energy and produce hard X-ray (HXR) bremsstrahlung. The deposited energy heats up the chromospheric plasma, causing it to expand upward into the loop where it becomes visible in soft X-ray (SXR) and extreme ultra-violet (EUV) emissions. This process has been termed chromospheric evaporation. It can be observed indirectly by comparison of SXR and HXR time evolution. If the observed SXR emission is a consequence of energy input by particle beams the time-integrated HXR emission should follow the SXR emission. This ``Neupert effect" \citep{Ne68} has been observed in many flares \citep[e.g.][]{De93,McT99,Ve05}. Direct observations of evaporating plasma were made early on in the form of upflows of hot plasma detected in SXR emission lines \citep[e.g.][]{1982ApJ...263..409A,1983SoPh...86...67A} and EUV \citep{Cz01, 2004ApJ...613..580B, Mi06b, Mi06a}.
Depending on the beam power it is generally distinguished between explosive and gentle evaporation. \citet{Fi85} showed that there is a threshold flux of 
$\sim$$10^{10}$ erg cm$^{-2}$s$^{-1}$ above which evaporation will be explosive, otherwise gentle. Explosive evaporation occurs on time-scales of $<$1 second. Due to large overpressure, heated plasma is driven up into the corona at up to several hundreds of km s$^{-1}$ \citep{1987ApJ...317..502F}, while upflows from gentle evaporation are generally slower \citep[order of tens of km s$^{-1}$ up to $\sim$200 km s$^{-1}$, e.g.][]{Za88,Pal83}. However, absolute velocities alone cannot be used for an unambiguous distinction between explosive and gentle evaporation since the velocities are temperature dependent, i.e. higher temperature plasma tends to have higher velocities. The overpressure during explosive evaporation causes a downward motion of the lower chromospheric layer \citep{Fi84,Fi85}. In explosive evaporation, lines that are formed at upper chromospheric and transition region temperatures (up to $\sim$1 MK) are thus red-shifted, while lines formed at higher temperatures are blue-shifted \citep[e.g.][]{1985ApJ...289..434F,2004ApJ...613..580B,Mi06b,2009ApJ...701.1209B}. In the case of gentle evaporation, both the lower temperature and the higher temperature lines will be blue-shifted \citep[e.g.][]{Za88,1988ApJ...333L..99Z,Mi06a,2009ApJ...701.1209B}. Observations of emission lines that form at different altitudes in the chromosphere are thus needed for unambiguous distinction. 

Chromospheric evaporation can also be triggered by a conductive energy flux in the absence of electron beams. This is the most likely mechanism leading to chromospheric evaporation in the decay phase of solar flares \citep[e.g.][]{An78,Za88,Cz01}, when a conductive flux is established between the hot post-flare loops and the cooler chromosphere. Conductively driven evaporation is also one explanation for the observations of increasing SXR emission and coronal density in the absence of HXR emission in the early phases of some flares \citep{Ba09}. 
\begin{figure*}
\resizebox{\textwidth}{!}{\includegraphics{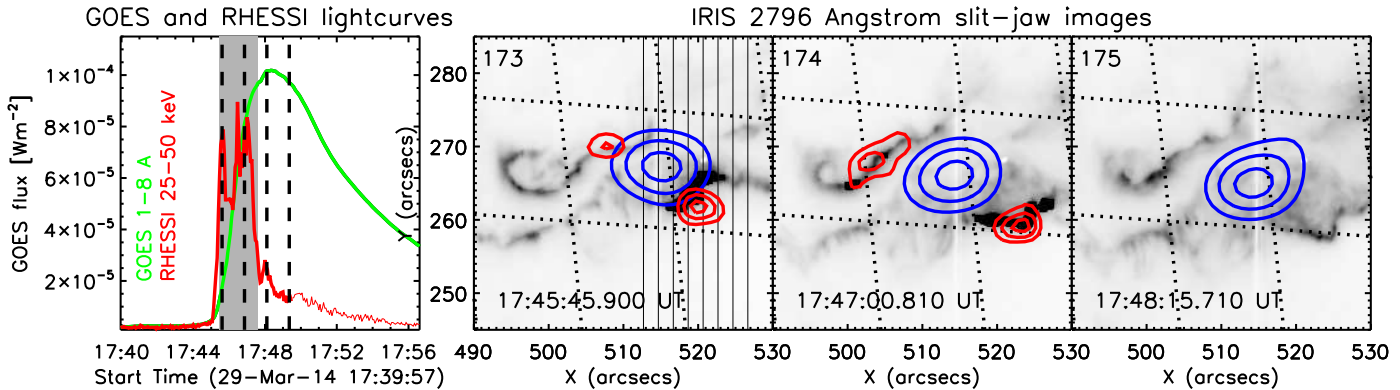}}
\caption{Left: GOES lightcurve (green) and RHESSI corrected count rate lightcurve (red, arbitrary units). The count rate after $\sim$17:48 UT was dominated by pulse pile-up (thin red line). The grey shaded area indicates the time during which HXR emission was imaged from the flare ribbons. The dashed lines mark the start times of the IRIS rasters analysed here (173 to 176).  Other panels: IRIS 2796 \AA\ slit-jaw images taken during rasters 173, 174, and 175 (times given on the maps) overlaid with 50\%, 70\%, and 90\% contours from RHESSI CLEAN images at 6-12 keV (blue) and 30-70 keV (red). The 8 slit-positions are indicated by vertical lines in the second panel.}
\label{fig1}
\end{figure*} 

For the case of electron beam driven evaporation, \citet{1987ApJ...317..502F} showed that there is a transition from explosive to gentle evaporation if the conductive flux out of the explosively heated plasma becomes comparable to the energy flux in the electron beam. This transition occurs on time-scales from a few seconds to a few tens of seconds. \citet{2009ApJ...701.1209B} present the analysis of such a transition in a flare observed by the Coronal Diagnostic Spectrometer (CDS) on onboard the Solar Heliospheric Observation \citep[SOHO,][]{Ha95}.  
More detailed studies of chromospheric evaporation became possible with the higher resolution and more complete temperature coverage of Hinode/EIS \citep{2007SoPh..243...19C}. \citet{2009ApJ...699..968M} and \citet{2010ApJ...719..213W} observed explosive evaporation in C-class flares with and without additional RHESSI \citep[Ramaty High Energy Solar Spectroscopic Imager,][]{Li02} observations. \citet{2011A&A...532A..27G} analysed a C-class flare using Hinode/EIS and RHESSI HXR observations. They find signatures of chromospheric evaporation but the RHESSI observations do not allow them to distinguish between electron-beam energy input or a purely thermal scenario.  \\

Here we present EUV observations of the \ion {Fe}{21} $\lambda$1354.1 line (formed at $\sim$10 MK, \citet{1970MNRAS.148...17J} and first observed in solar flares by \citet{1975ApJ...196L..83D}). Combined observations were made with the Interface Region Imaging Spectrograph \citep[IRIS,][]{2014SoPh..289.2733D}, RHESSI, and EIS during the X-flare on 2014 March 29. The high spatial resolution of IRIS allows for tracing the location, velocity, and timing of hot evaporated plasma along the flare ribbon. Combining these observations with RHESSI imaging and spectroscopy, it is possible to investigate the type of energy input that causes evaporation (electron-beam versus conductive energy input) and its nature (explosive versus gentle). 
\section{OBSERVATIONS AND DATA ANALYSIS} \label{data analysis}
The GOES X1 flare occurred on 2014 March 29 with the HXR rise starting at 17:44 UT and peaking at 17:47 UT. RHESSI imaging and spectroscopy give the timing, location and amount of electron-beam deposited energy. 

IRIS observations of the \ion {Fe}{21} $\lambda$1354.1 line give the location, speed, and intensity of evaporating material at a temperature of $\sim$10 MK. 
The IRIS data were obtained during a coordinated flare observing campaign on 2014 March 29 with the observation lasting from 14:09-17:54 UT \citep{kleint15}. A total of 180 8-step rasters with a cadence of 75 s were performed. In the paper we refer to the raster numbers of the Level 2 filenames (starting at raster 000 to raster 179), for which all relevant calibrations, such as flat-fielding, dark-correction, geometric correction, and wavelength correction have already been applied \citep{2014SoPh..289.2733D}. Each raster had a field-of-view of 14$\arcsec$ x 174$\arcsec$. During the X1 flare, the western ribbon and a small part of the eastern ribbon were caught during several rasters (compare Figure \ref{fig1}). The far UV (FUV, 1332--1358/1389--1407 \AA) data had an exposure time of 8 s and a dispersion of 25.46 m\AA\ pixel$^{-1}$ with a plate scale of 0.166$\arcsec$ pixel$^{-1}$. The \ion {Fe}{21} spectral window, which we analyze here, was not overexposed for the whole duration of the observations and at its blue-ward edge is limited to 350 km s$^{-1}$ from the line center at rest (compare Figure \ref{irisl}). In the flare presented we find that the entire \ion {Fe}{21} line profile is shifted but always lies completely within this window.
Here we focus on the main HXR peak when HXR emission was observed from the flare ribbons (grey area in Figure \ref{fig1}) and where considerable upflows were detected in \ion {Fe}{21}. Figure \ref{fig1} shows SXR lightcurves from GOES and HXR RHESSI lightcurves. The start times of the analyzed IRIS rasters (rasters 173, 174, 175, 176) are indicated.
\subsection{X-ray analysis}\label{xrayan}
RHESSI full-sun spectra were fitted during the time of the main peak, using 8 second time intervals coinciding with the start and end times of the IRIS exposures at each slit position. The spectra were fitted with a thermal component at low energies and a thick-target power-law component at energies above $\sim$20 keV. 
RHESSI CLEAN \citep{Hur02} images at 6-12 keV (using grids 3-6, natural weighting with a clean beam width factor of 1.4, resulting in an effective CLEAN beam of 12.8$\arcsec$) and at 30-70 keV (using grids 1-6, natural weighting with a clean beam width factor 1.4, resulting in an effective resolution of 3.4$\arcsec$) were made for an overview of the flare morphology. We integrated over 75 seconds from the start time of each IRIS raster to be consistent with the time it takes to complete one raster (except for the first raster where the RHESSI attenuator came in during the period hence the start time of the image was 17:46:02 UT). For the more detailed analysis of the time evolution, additional RHESSI images were made with the same start times as the IRIS exposures at each slit position and an 8 seconds time-integration. To ensure as close a representation of the true source size as possible, only grids 1--4 were used for these images with an effective resolution of 3 arcsec FWHM. These images were used in the further analysis. The RHESSI X-ray images show a SXR source near the top of a loop system that connects the two flare ribbons. The HXR footpoints coincide with the location of the flare ribbons as seen in the IRIS 2796 \AA\ slit-jaw images (SJI) but are less extended than the flare ribbon along the east-west direction and do not cover its entire length. Figure \ref{fig1} shows IRIS 2796 \AA\ slit-jaw images at selected times overlaid with the RHESSI contours at 6-12 keV and 30-70 keV. We found a $\sim$3 \arcsec  offset between the RHESSI sources and SDO images of the flare ribbons, to which IRIS was aligned. Thus we applied an empirical roll correction of 0.15 degrees (clockwise about Sun center) to the RHESSI images, so that the HXR emission matches emission from both flare ribbons in SDO \citep[see also][]{kleint15}. This correction is justified for several reasons. In addition to the pointing uncertainty of SDO, the star-field that is used by the PMTRAS \citep[Photo-Multiplier Tube Roll Aspect System,][]{2002SoPh..210..101H}to determine the RHESSI roll-angle was sparse at the time of the observations, resulting in a potential error of the RHESSI roll-angle. Moreover, the correction is justified because of physical reasons. With no correction applied, the HXR footpoints would not be co-spatial with the flare ribbons as seen in IRIS slit-jaw images and they would be trailing the motion of the ribbon. In addition, the HXR sources would be located in the same magnetic polarity. All these points would be in contradiction of our understanding of flares and flare energy input. 
\subsection{FUV analysis}
\begin{figure}[]
\includegraphics[width=8cm]{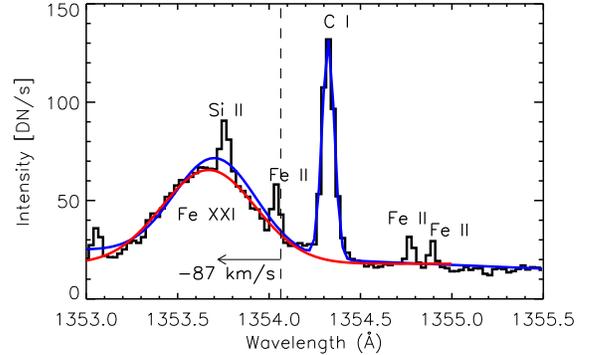} 
\caption{Example of fits to the \ion {Fe}{21} $\lambda$1354.1 line to infer the Doppler-velocity using two different methods (red: single Gaussian fit, blue: double-Gaussian fit). The dashed line marks the rest wavelength of \ion {Fe}{21} $\lambda$1354.1.}
\label{fig3}
\end{figure}

IRIS observations of \ion {Fe}{21} $\lambda$1354.1 allow for studying flows of hot ($\sim$10 MK) plasma near the loop footpoints. IRIS observations of \ion {Fe}{21} were recently reported in selected events \citep{2014ApJ...797L..14T, polito15,2015ApJ...807L..22G}, including the 2014 March 29 flare by \citet{young14}, who observed blue-shifts from the flare ribbons ``as expected from models of chromospheric evaporation" without going into quantitative detail. 
To find the center location of the \ion {Fe}{21} line, we fitted two Gaussians, one to the \ion {Fe}{21} line, the other to account for the intense \ion{C}{1} $\lambda$1354.29 line. The fits were done automatically over time and along the slit. A rest-wavelength of 1354.064 \AA, following \citet{2000ApJ...544..508F}, was used to calculate the Doppler-velocities. Fits that resulted in a line-width smaller than the theoretical thermal line-width, with too large $\chi ^2$, as well as with negative or unphysically high intensities were omitted from further analysis. As validation of the fits we also fitted a single Gaussian to the \ion {Fe}{21} line, omitting contributions from \ion {C}{1} and other, weaker lines. Both methods give largely consistent results (to within a few km/s), indicating that the weaker lines (originating from cool, singly ionized Fe and Si) are not important for the fit and the two-component Gaussian fit, which we use for inferring the Doppler-velocity, is justified. An example is shown in Figure \ref{fig3}. It shows the broad, blue-shifted  \ion {Fe}{21} line centered 1353.6 \AA, the intense, narrow \ion {C}{1} line at 1354.35 \AA,
and the weaker \ion {Fe}{2} and \ion {Si}{2} lines \citep[see][for a more detailed discussion of the whole spectrum]{2015arXiv150502736T}.
\section{Results}
HXR footpoints can be imaged starting at 17:45:28 UT until 17:47:25 UT (compare Figure \ref{fig1}). Emission above $\sim$20 keV is visible in lightcurves and spectra after this time but becomes increasingly dominated by pile-up \citep{Smith02} after $\sim$17:49:30 UT. The RHESSI spectral fits indicate a hot coronal source with a temperature of 25 MK at the start of the analysis interval. The temperature drops to 21 MK at 17:49 UT. The total electron number flux per second above 20~keV decreases from $2\times 10^{35}$ to $4\times 10^{35}$ though the electron spectral index remains constant overall with $\delta \approx 3.5$.

\begin{figure}
\includegraphics[width=8cm]{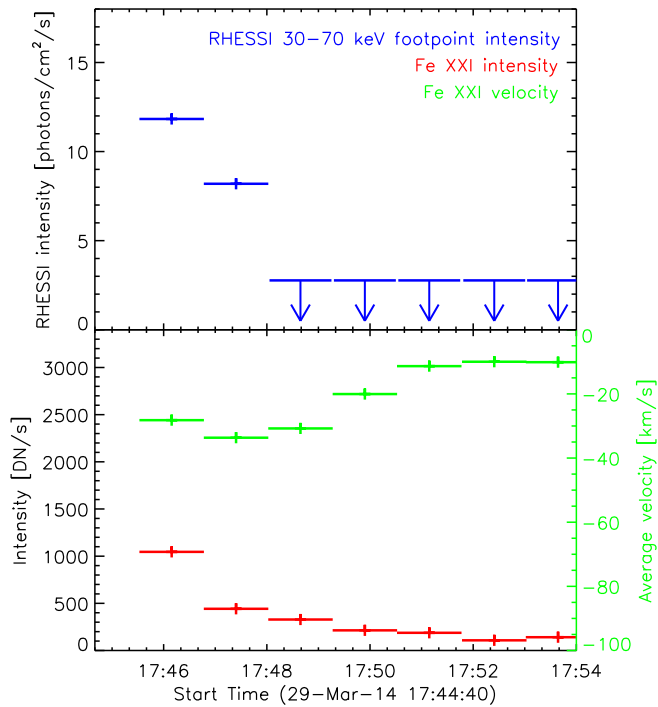} 
\caption{Top: Spatially integrated HXR footpoint intensity of 30-70 keV emission for the same time steps as IRIS. The arrows indicate a conservative upper limit of the HXR emission for the time-steps where no imaging was possible due to limited statistics, indicating that upflows were present several minutes after the end of the detectable HXR emission. Bottom: Spatially integrated intensity of blue-shifted \ion {Fe}{21} lines, time-averaged over a whole raster (red points) and average upflow velocities (green points).}
\label{figtvol}
\end{figure}

\begin{figure*}[]
\resizebox{0.9\textwidth}{!}{\includegraphics{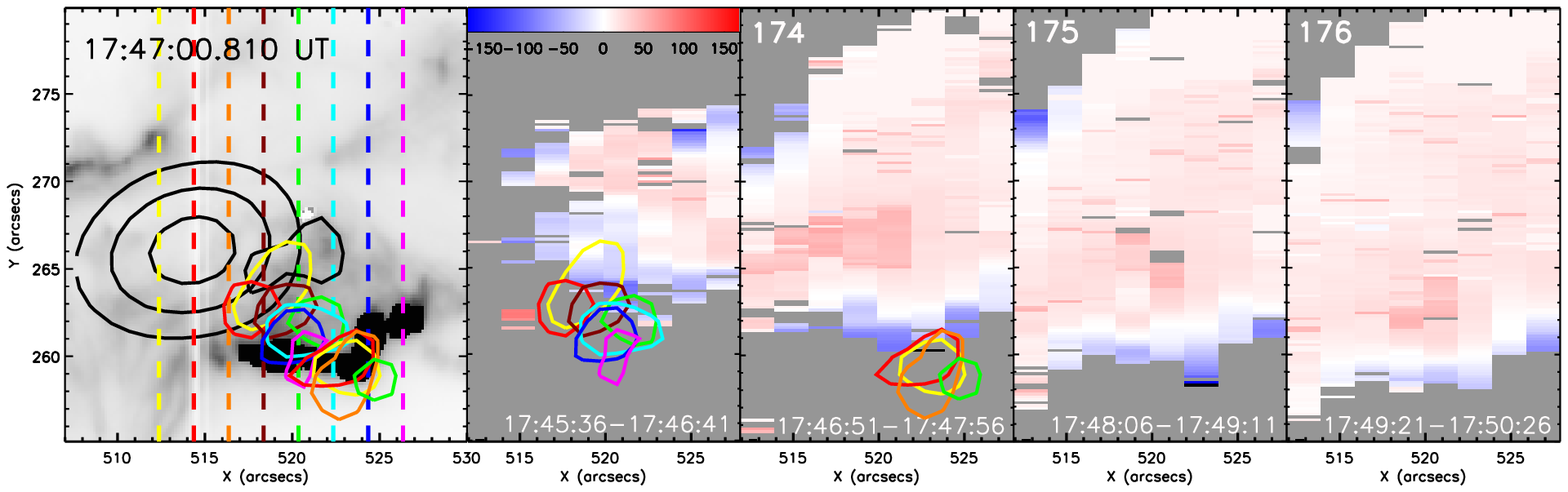}} \\
\resizebox{0.92\textwidth}{!}{\includegraphics{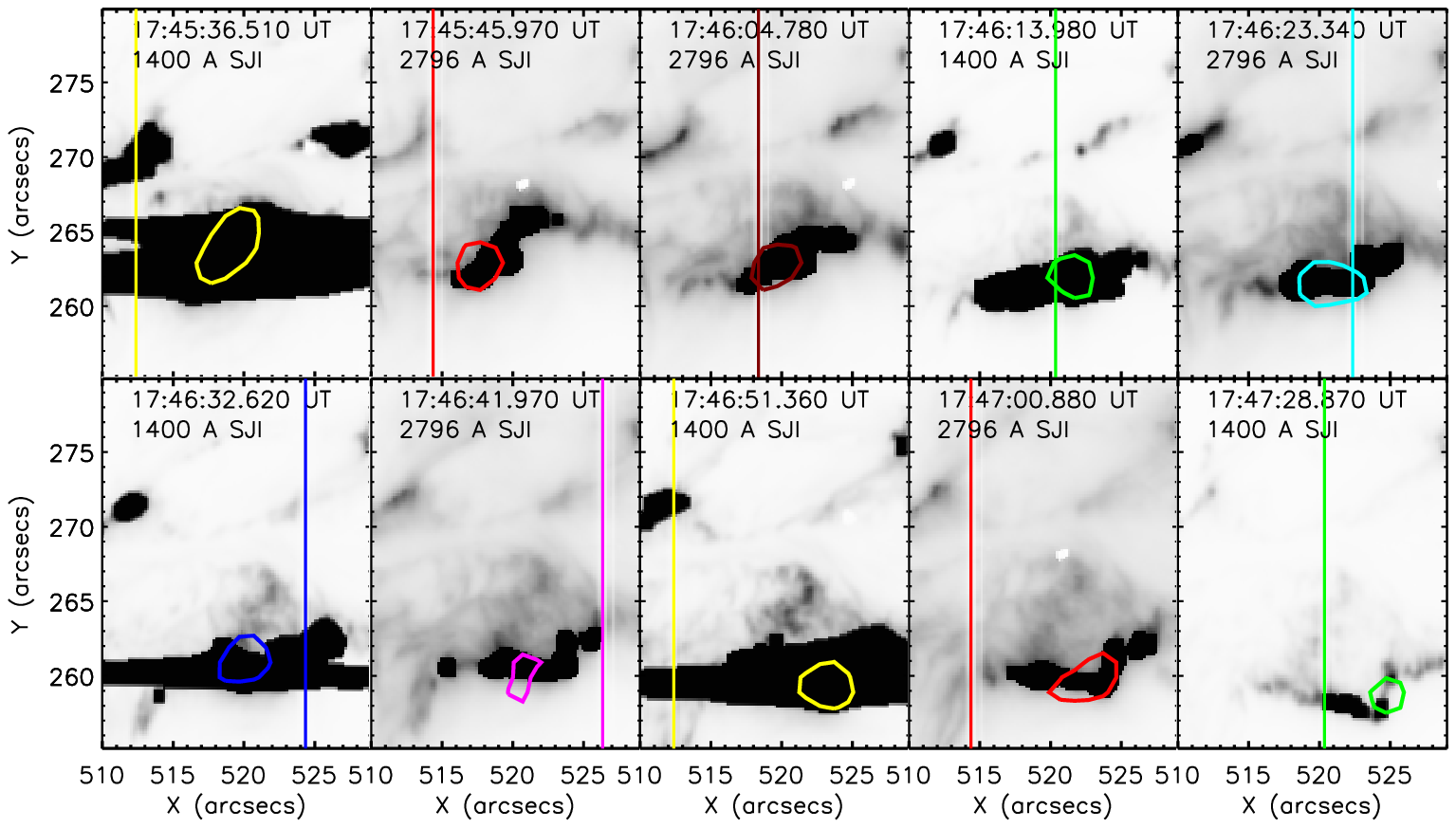}}
\caption{Top left: IRIS 2796\AA\ slit-jaw image overlaid with 50\%, 70\%, and 90\% contours from RHESSI CLEAN images at 6-12 keV (black) and 30-70 keV (colors, 70\% contours only). The slit positions are indicated with coloured dashed lines. The RHESSI contours show the position of the HXR footpoint for each slit-position in the respective color. Top, other panels: Doppler-velocity maps of \ion {Fe}{21} along the slit for rasters 173 to 176 (indicated by white numbers). Grey areas denote pixels where either no \ion {Fe}{21} emission was detected or the fit was bad. The contours of the HXR footpoints observed during a given raster are overlaid in the respective colors of the slit-positions. Middle and bottom row (from left to right): Individual slit positions relative to HXR footpoint location during rasters no. 173 and 174, overlaid on IRIS  2796 \AA\ and 1400 \AA\ slit-jaw images (raster steps where no HXR footpoints were observed or no 2796 \AA\ and 1400 \AA\ images were available are omitted).}
\label{figblue}
\end{figure*}

The intensity enhancement in \ion {Fe}{21} was first observed during the raster that started at 17:45:36 UT (number 173) and lasted well into the decay phase of the flare. Comparison of the time evolution between HXR emission and blue-shifted \ion {Fe}{21} emission is difficult to illustrate since one IRIS raster takes 75 seconds to complete and the HXR footpoint moves about 7 $\arcsec$ over the course of the observations. As an overview we calculate the intensity in blue-shifted lines as a function of time, spatially integrated over each raster. The total HXR footpoint intensity of 30--70 keV emission stronger than 50\% of the maximum in each image was computed for comparison. The resulting lightcurves are displayed in Figure~\ref{figtvol}. The HXR footpoint emission was most intense during raster 173, when the \ion {Fe}{21} intensity was also the most intense. However, HXR emission was not observed from raster 175 onward, while blue-shifts persisted until the end of the IRIS observations. Here we focus on the four rasters where the intensity enhancement in \ion {Fe}{21} was most pronounced (173-176) and investigate the temporal and spatial association of \ion {Fe}{21} blue-shifts with HXR footpoints in detail to identify the trigger of chromospheric evaporation.
Figure \ref{figblue} shows the full time evolution of HXR source locations relative to the IRIS slit-position. The IRIS slit positions are color-coded and the contours of the HXR source are plotted in the respective colours. For example, for raster 174, taken between 17:46:51-17:47:56 UT, HXR footpoints could only be imaged for slit-positions 1-3 and 5 (yellow to green contours). The HXR footpoints are indicated by 70\% contours of the maximum emission. Since the sources are likely not resolved, this contour level provides an upper limit on the area of maximum energy deposition.
Three cases can be distinguished: 1) regions with strong blue-shifts in \ion {Fe}{21} that are associated with the presence of HXR footpoints in space, but not time, i.e. the HXR footpoint was observed several tens of seconds earlier; 2) regions where the IRIS slit was co-spatial with the location of the HXR source so that  blue-shifts were observed co-temporally with the HXR footpoints, yet, interestingly, not from the same location; 3) regions with blue-shifts in \ion {Fe}{21} that have no association with HXR emission in space or time. An example for each case is shown in Figure \ref{thefigure}. 
\subsection{Spatial Association of HXR footpoint emission with \ion {Fe}{21} blue-shifts}\label{cospat}
An example for this case is shown in panel one of Figure \ref{thefigure} for slit position 5 of raster 174. The blue-shifts were observed at 17:47:28 UT. The HXR source at the same location was observed $\sim$56 s earlier, at 17:46:32 UT. Near the time of the IRIS observation, the HXR footpoint was $\sim$5 arcsec away to the south-west. For other slit positions, HXR footpoints preceded the appearance of blue-shifts at a given slit-position by 30-60 seconds on average.
The velocities in all such regions were between -30 km s$^{-1}$ to -100 km s$^{-1}$. Corrected for line-of sight, this gives a maximum velocity of $\sim$-120 km s$^{-1}$.  The location of the most pronounced upflows is co-spatial with the location of the ribbons in slit-jaw images at 2796\AA\  supporting the notion that the observed blue-shifts are indeed due to chromospheric evaporation of 10 MK plasma. 
\subsection{Co-temporal observation of HXR emission and upflows} \label{cotemp}
During raster 173 the slit crossed the HXR footpoint at positions 4 and 5. An example is given in the second panel of Figure \ref{thefigure}. The HXR footpoint is located at $\sim$[519,262.5] arcsec. In a beam-driven evaporation model, blue-shifts are expected from that same location. However upflows are observed 2-3 arcseconds further north but not from the location of the HXR footpoint. 
\subsection{Upflows without association with HXR emission} \label{nohxr}
\begin{figure*}[]
\resizebox{\textwidth}{!}{\includegraphics{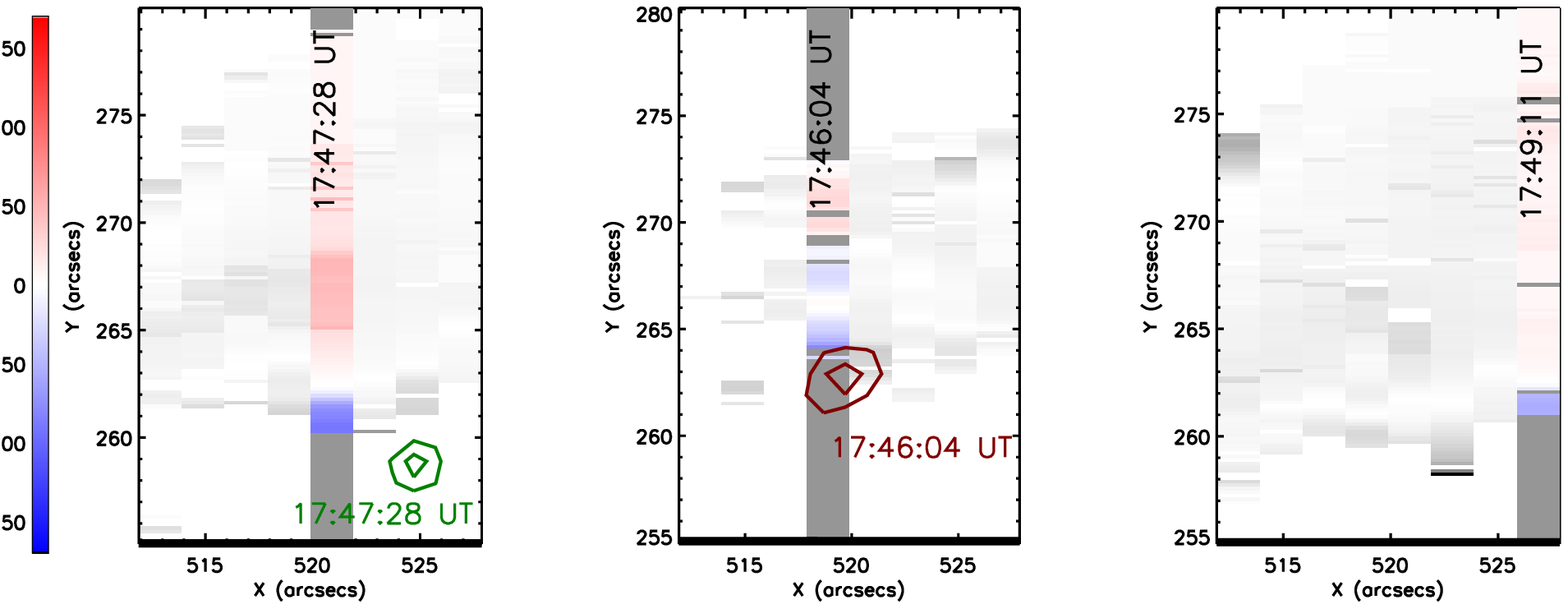}}
\caption{Examples of relative location and timing between HXR footpoints and blue-shifts in \ion {Fe}{21}. Left:  blue-shifts observed $\sim$56 seconds after the HXRs (see Section 3.1) at slit position 5. The observations of blue-shifts at about [521,260] arcsec were made at 17:47:28 UT. The co-temporal HXR source was at $\sim$[525,259] arcsec (green). 70\% and 90\% contours are given for the HXR sources. Middle: co-temporal observations (Section 3.2). The IRIS slit was co-spatial with the HXR footpoint, but the observed upflows originate from a location where the HXR source was observed 28 seconds earlier. Right: upflows that have no association with HXR emission in space or time (Section 3.3).}
\label{thefigure}
\end{figure*}

There are regions that display blue-shifts in \ion {Fe}{21} but no corresponding RHESSI HXR source at any time, namely at slit-positions 7 and 8, and the northern ribbon (compare with Figure \ref{figblue}). An example is given in the third panel of Figure \ref{thefigure}. The velocities are similar to the velocities at the other locations with a maximum upflow velocity of -110 km s$^{-1}$ (-132 km s$^{-1}$ line-of-sight corrected) in the northern ribbon and up to -200 km s$^{-1}$ in the southern ribbon. 
\section{Discussion and Conclusions}
The observations show evidence of chromospheric evaporation of plasma at a temperature of 10 MK from both flare ribbons. In a standard scenario, chromospheric evaporation is the consequence of energy deposition by a non-thermal electron beam. The timing of observed upflows in \ion {Fe}{21}, their location, and the measured velocities all point towards gentle evaporation in the presented case, yet the HXR observations suggest a large energy input by a non-thermal beam. In the following we investigate whether the non-thermal energy deposition is large enough to trigger explosive evaporation, followed by a discussion of the three cases presented above and how evaporation is produced in each case.
\subsection{Energy deposition by non-thermal particle beams}
The total non-thermal power in a beam of accelerated electrons with flux distribution F(E) is given as 
\begin{equation}
P_{tot}=\int E F(E) dE.
\end{equation}
For a power-law spectrum $F(E)=AE^{-\delta}$ above a cut-off energy $E_{c}$ this gives
\begin{equation}
P_{tot}=\frac{A}{\delta-2}E_c^{-(\delta-2)}\,\mathrm{erg\,s^{-1}}
\end{equation}
or, expressed through the total electron flux per second $F_{tot}$ found from a thick-target model fit:
\begin{equation}
P_{tot}=\frac{\delta-1}{\delta-2}F_{tot}E_c \,\mathrm{erg\,s^{-1}}
\end{equation}
For the time-step starting at 17:46:04 UT, the total electron flux from the thick-target fit was $2.3\times 10^{35}$ $s^{-1}$, electron spectral index $\delta=3.9$, and cut-off energy $E_{c}=20$ keV. This gives a total non-thermal power of $P_{tot}=1.1\times 10^{28}$ erg s$^{-1}$. At 17:47:10 UT, the non-thermal power had dropped to  $P_{tot}=5.5\times 10^{27}$ erg s$^{-1}$. Since the fits were made on full-sun spectra, these values include emission from both footpoints. RHESSI images suggest that the southern footpoint was more intense than the northern footpoint during most of the flare. Assuming that they were of roughly equal intensity gives a lower limit on the electron flux into the southern footpoint of $P_{tot}=5.5\times 10^{27}$ erg s$^{-1}$ and $P_{tot}=2.8\times 10^{27}$ erg s$^{-1}$, respectively. The footpoint area, estimated from IRIS 2796 \AA\ images is $\approx 10^{17}$ cm$^2$. A similar value is found from the 50\% contour in RHESSI CLEAN images. This is an upper limit, since the IRIS images were partially saturated and source sizes with RHESSI tend to be over-estimated \citep[e.g.][]{2013A&A...552A..86W,De09}. However, this estimate is sufficient here, since we are interested in a lower limit of the energy input. The total energy flux amounts are $5.5\times 10^{10}$ erg s$^{-1}$cm$^{-2}$ initially and $2.8\times 10^{10}$ erg s$^{-1}$cm$^{-2}$ by 17:47:10 UT.  
\subsection{Spatial Association of HXR footpoint emission with \ion {Fe}{21} blue-shifts}\label{cospatdisc}
According to the above calculation, the non-thermal power input is big enough to trigger explosive evaporation.  According to \citet{1987ApJ...317..502F}, explosive evaporation ceases and becomes gradual once the conductive flux out of the evaporated plasma becomes comparable to the beam flux. The change occurs at temperatures around $\sim$10 MK over time-scales of a few seconds to a few tens of seconds. In the presented event, the time scale would be of the order of 10 seconds. This scenario could explain blue-shifts that are still observed long after the HXR source since gentle evaporation will continue as long as there is a temperature gradient. Gentle evaporation is further supported by the observed velocities of less than 200 km s$^{-1}$. Thus the observations described in Section \ref{cospat} can be explained with energy input by a non-thermal electron beam resulting in explosive evaporation followed by a transition to gentle evaporation whose signatures are observed once the IRIS slit covers the respective area. 
\subsection{Co-temporal observation of HXR emission and upflows} \label{sec-cotemp}
If energy input by electron beams is indeed the initial cause of the evaporation, as assumed in the scenario discussed above, the question immediately arises: Why are no blue-shifts observed when the IRIS slit is co-spatial with the location of the HXR footpoint? In other words, why is there no high temperature signature of evaporation for the case described in Section \ref{cotemp}? One potential explanation could be the delayed onset of EUV emission due to the ion equilibration time. Heating by beam-energy input is almost instantaneous. However, when neutral Fe is heated quickly from 10000 K to 10 MK it takes a certain time to reach ionisation equilibrium, depending on the ambient density. \citet{2009A&A...502..409B} showed that for densities $>10^{12}$ cm$^{-3}$ this time-scale is shorter than one second. For densities of $10^{10}$ cm$^{-3}$ equilibration takes of the order of 100s.  In a standard thick target, assuming an exponential density, the bulk of 20 keV electrons will reach heights between 1000 km and 1400 km, corresponding to densities between 1$0^{12}$ cm$^{-3}$ to $5\times10^{13}$ cm$^{-3}$  \citep{Ba11a,Br02}  thus this effect can be ignored and one should expect signatures of evaporation in an 8 second integrated spectrum. However, it is likely that the \ion {Fe}{21} emission is obscured by chromospheric lines in the spectral window. As shown by \citet{2015ApJ...807L..22G}, the earliest, and most shifted, instances of the \ion {Fe}{21}  line can be extremely weak at the footpoints. Due to the small spatial scales investigated here, accurate co-alignment between instruments is crucial. For the reasons explained in Section \ref{xrayan} we are confident that the alignment adopting a minimum 0.15 degree rotation is correct. If no such corrections were applied, the HXR footpoint would cover the southern edge of the blue-shifted area in Figure \ref{thefigure} (middle) but would still not cover it fully. Thus an error in co-alignment alone cannot account for the observed offset.
\subsection{Upflows without association with HXR emission}
Chromospheric evaporation in the absence of HXR emission can most readily be attributed to energy input by thermal conduction from a hot coronal source. The conductive energy input from classical Spitzer conductivity \citep{Spbook} is given as
\begin{equation}
L_{cond}=10^{-6}T^{5/2}\nabla T \,\mathrm{erg\,s^{-1} cm^{-2}}.
\end{equation}
This is usually approximated to 
\begin{equation}
L_{cond}=10^{-6}\frac{T^{7/2}}{L_T}
\end{equation}
with $L_T$ the temperature scale-length. The RHESSI spectra indicate a hot ($\sim$25 MK) coronal source that persists during much of the decay phase. The temperature scale-length is the loop-half length which is approximated from the footpoint separation, assuming a semi-circular loop. For a footpoint separation of 21.4 arcsec this gives $L_{T}\approx 1.2\times 10^9$ cm. The resulting conductive flux is then $L_{cond}\approx 2.9\times 10^{9}\,\mathrm{erg\,cm^{-2}s^{-1}}$. In typical flare conditions, the heat flux is expected to saturate \citep{Ba09}. \citet{Ca84} showed that this can be accounted for by a reduction factor that only depends on the electron mean free path and the temperature scale length. In the present case, this factor amounts to $\approx 0.85$ resulting in a reduced conductive flux of $\approx 2.2\times 10^{9}\,\mathrm{erg\,cm^{-2}s^{-1}}$ as a lower limit. Such a conductive energy input would be sufficient to drive the observed evaporation. 
Another possibility are electron beams whose signatures are not observed due to RHESSI dynamic range (about 10:1). A source whose intensity at a given energy is less than one tenth of the peak intensity cannot be imaged. Assuming an electron beam with the same low-energy cutoff and spectral index as used above but ten times less flux would result in a total power input reduced by about one order of magnitude, enough to lead to gentle evaporation.  
\subsection{Other flow velocities in IRIS and EIS}
\begin{figure*}
\includegraphics[height=10cm]{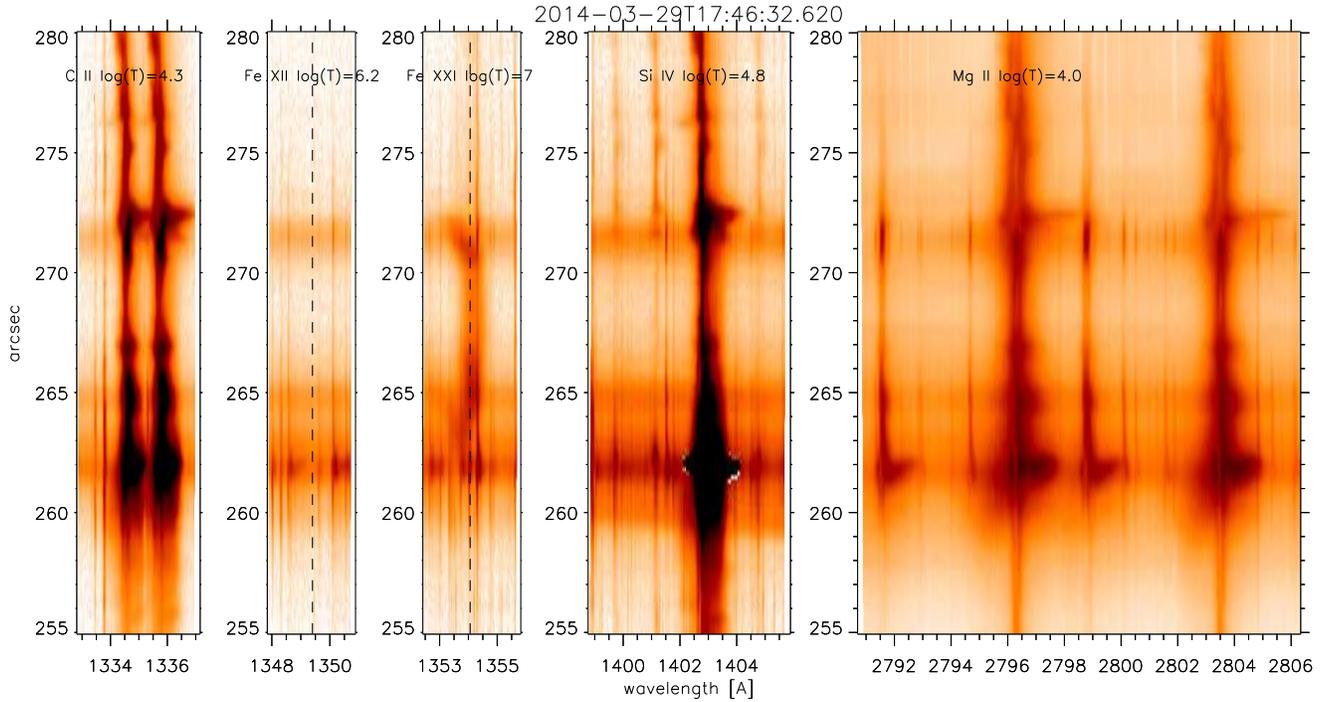} 
\caption{A selection of spectral regions from IRIS (from left to right): the two \ion{C}{2} lines; \ion{Fe}{12} which is generally faint and not visible here (rest wavelength indicated by dashed line); \ion{Fe}{21} which is well visible as inverse C-shape; several \ion{O}{4} lines (1399.8 \AA, 1401.2 \AA, 1404.8 \AA) and the strong \ion{Si}{4} line at 1402.8 \AA; part of the near-UV (NUV) window with the Mg k and h lines (2796.4 \AA, 2803.5 \AA). The colors are inverted and the images are contrast-enhanced. Downflows in \ion{Si}{4}, \ion{Mg}{2} and \ion{C}{2} are seen just north and south of the maximum blueshifts of \ion{Fe}{21} (Y $\approx$ 272\arcsec\ and 263\arcsec).}
\label{irisl}
\end{figure*}
An unambiguous distinction between gentle and explosive evaporation is not possible based on the observation of the velocities at one temperature alone, but could be made via doppler analysis of, in particular, the O IV line which forms at T$\sim$0.16 MK. For gentle evaporation one would expect to see blueshifts in the O IV line \citep{2004ApJ...613..580B}, while redshifts are expected for explosive evaporation. However, the data from IRIS are inconclusive showing very low velocities (10 km s$^{-1}$) in the O IV line, as well as mixed up- and downflows at locations where evaporation is seen in \ion {Fe}{21}. Figure~\ref{irisl} shows a selection of five spectral windows taken by IRIS at 17:46:32 UT (solar X $\approx$ 524 arcsec). The third panel shows the rather broad \ion{Fe}{21} emission with its rest wavelength 1354.06 \AA\ indicated by the dashed line. The central part (Y $\approx$ 265 - 270 arcsec) does not show significant \ion {Fe}{21} velocities, only a small redshift. Northward of Y $\approx$ 270 and southward of Y $\approx$ 265 \ion{Fe}{21} shows a clear blueshift on the order of 0.5 \AA, corresponding to $\approx$ 110 km s$^{-1}$. The other spectral lines do not show clear shifts at these locations, rather some line-broadening. However, Si IV, Mg II and C II show prominent downflows just north and south of these locations, which correspond to leading edges of flare ribbons. It is possible that \ion{Fe}{21} has not formed yet there, which would indicate densities of the order of 10$^{10}$ cm$^{-3}$ according to the equilibration time (see Section \ref{sec-cotemp}). Or, \ion{Fe}{21} could simply be too faint to be visible at these locations. 
The flare was also observed by Hinode/EIS and twice ($\sim$17:46:28 UT and $\sim$17:48:52 UT) the EIS slit was at the same location as the IRIS slit. The line selection was rather sparse and does not provide full temperature coverage. As expected, the upflow velocities in EIS \ion {Fe}{23} (12.5 MK) are higher than those seen in \ion {Fe}{21} in IRIS. However, analysis of the other available data suggests red-shifts in Fe XVI (2.8 MK) and potentially in Fe XVII (5.6 MK) at both observed times. It is interesting that there should be down-flows at temperatures much higher than found by \citet{2009ApJ...699..968M}. For the time at 17:48:52 UT this would indicate that evaporation is explosive even long after particle acceleration has ceased but no definite statement can be made from the available data.\\

In summary, for parts of the presented flare, the timing, location, and velocities of the observed signatures of chromospheric evaporation with IRIS and RHESSI can be explained with conductive energy input due to the temperature gradient between the chromosphere and evaporated, 10 MK plasma in the loop, as well as the hot coronal SXR source. It is intriguing however, that upflows are not observed from the same location as the HXR sources where the IRIS slit was co-spatial with the HXR source. On the other hand, the EIS data could be interpreted as showing explosive evaporation even at times when no HXR emission was observed.

Future combined high-spatial resolution flare observations, including lower temperature lines such as \ion{O}{4} and \ion{Si}{4} and in combination with Hinode/EIS should help shed some light on the matter.

\mdseries
\acknowledgments
This work was supported by the Swiss National Science Foundation (200021-140308) and through NASA contract NAS 5-98033 for RHESSI. LK was supported by a Marie Curie Fellowship. DG acknowledges support by the European Community's Seventh Framework Programme (FP7/2007-2013) under grant agreement no. 606862 (F-CHROMA). We would like to acknowledge the International Space Science Institute (ISSI) for facilitating the collaboration on this work. 

\bibliographystyle{aa}
\bibliography{../../mybib}

\end{document}